\documentclass{article}
\usepackage{amssymb,amsfonts,amsthm,amsbsy}
\usepackage{graphics}
\usepackage{graphicx}
\newcommand{\be}{\begin{equation}}
\newcommand{\ee}{\end{equation}}
\newcommand{\bea}{\begin{eqnarray}}
\newcommand{\eea}{\end{eqnarray}}

\newcommand{\bc}{\begin{center}}
\newcommand{\ec}{\end{center}}

\newcommand{\beq}{\begin{eqnarray*}}
\newcommand{\eeq}{\end{eqnarray*}}

\def\bi{\bibitem}

\def\hsp5{\hspace{5mm}}

\def\case#1/#2{\textstyle\frac{#1}{#2}}

\begin{document}
\title{\sc Serial Correlation, Periodicity and Scaling of Eigenmodes in an Emerging Market \\ }
\author{{\sc Diane Wilcox$^{\dag}$ $^{\ddag}$ \ \& \ Tim Gebbie$^{\dag}$ }  \\ \\
 $^{\dag}${\it Department of Mathematics \& Applied Mathematics}, \\
{\it University of Cape Town, Rondebosch, 7700, South Africa} \\ \\
$^{\ddag}${\it Brait Specialised Funds, Postnet Suite 80,} \\ {\it Private Bag X1005, Claremont, 7735, South Africa} \\ \\
\small  diane.wilcox@uct.ac.za, \ \
\small tim.gebbie@physics.org }
\date{2 September 2007}
\maketitle

\begin{abstract}
We investigate serial correlation, periodic, aperiodic and scaling behaviour of
eigenmodes, i.e. daily price fluctuation time-series derived from
eigenvectors, of  correlation matrices  of shares listed on the
Johannesburg Stock Exchange (JSE) from January 1993 to December
2002.

Periodic, or calendar, components are detected by
spectral analysis. We find that calendar effects are
limited to eigenmodes which correspond to eigenvalues outside the
Wishart range. Using a variance ratio test,
we uncover serial correlation in the first eigenmodes
and find slight negative serial correlation for eigenmodes within the Wishart range. Our spectral analysis and variance ratio investigations suggest that interpolating missing data or illiquid trading days with zero-order hold introduces high frequency noise and spurious serial correlation.
  Aperiodic and scaling behaviour of the eigenmodes are investigated
by using rescaled-range (R/S) methods and detrended fluctuation analysis
(DFA). We find that DFA and classic and modified R/S exponents
suggest the presence of long-term memory effects in the first five eigenmodes.
\end{abstract}

\vspace{0.3cm}



%
%
%

\section{Introduction}

Any complex system of interacting variables that generates
stochastic time-series data may be inspected via its covariance
matrix properties. In finance the portfolio theory of Markowitz
\cite{markowitz} is premised on complete knowledge of the full
covariance matrix of assets in the investment universe. In
practice this is not possible for financial time-series data
because: (1) incomplete and limited amounts of data are available,
(2) measurement errors are ubiquitous, and (3) structural
variations in processes generating data lead to estimated
covariances that are rarely stable in time. This necessitates a
methodology for reducing the dimension of the problem.

The arbitrage pricing theory of Ross \cite{Ross76} explains
equilibrium stock returns by a multitude of risk factors. Stocks
with the same risk factor loadings should provide the same
equilibrium returns. This is the arbitrage free notion implied by
the theories name. By providing a framework for separating signal
from noise and systematic risk from stock specific risk, the
theory gives a means for the aggregation of stock price data for
modelling and understanding individual stock returns
\cite{RollRoss80}. The arbitrage pricing theory does provide a
theoretical framework for dimensional reduction; it does not
specify how many factors to use.

Identifying risks that are shared across the market and which are
not unique to individual stocks is central to understanding the
collective behaviour of stock prices and the stock market as a
whole{\footnote{Stock market indices are inadequate in this regard
because index construction typically focusses on a limited number
of factors as the basis of construction. Such a limited basis does
not span the market and merely provides a projection thereof.}}.
These risk factors, the timeseries constructed from eigenvector
components or {\em eigenmodes} (Equation (\ref{emodes})) of the
estimated covariance matrix, are of the focus of this paper.
Extracting the orthogonal modes is a method of de-correlating
time-series and serves as method of dimensional reduction.



In \cite{WiGe1,WiGe2} we applied RMT methods to investigate how
the treatments of missing data and thin trading (no price changes
recorded for a stock over several time periods) impact on the
computation of cross-correlations in an emerging market. Several studies have applied random matrix theory
(RMT)\footnote{In particular, known universal properties for
Wishart matrices such as the Wishart distribution for eigenvalues
\cite{bai,boupot,pre60a}, the Wigner surmise for eigenvalue
spacings \cite{brody,guhr,mehta2,mehta}, the Porter-Thomas
distribution of and the inverse participation ratio (IPR) for
eigenvector components \cite{guhr,prl83a,pre65a}, amongst others.}
to calibrate and reduce the effects of noise in cross-correlation
matrices
\cite{BoLiMa1,DrKwGrRuSp1,drKwSpWo1,physa259,pre64a,prl83a,euphysj11a,manstan,prl83b,pre65a}. Our
investigation was based on 10 years of daily data for traded
shares listed on the JSE Main Board from January 1993 to December
2002. The data set used incorporated a zero-order hold for prices
when there was no trading. This accounts for sequences of
zero-valued returns in the return times-series even though no
measurements occurred. The two estimators were aimed at exposing
the effects of thin trading or missing data on the resulting
covariance matrices in the context of RMT. For all our estimated
covariance matrices, we found that the eigenvalue distributions
exhibited the following: (1) a significant part of the spectrum
fell within the range of random matrix predictions,(2) there were
a small number of large leading eigenvalues, and (3) there was a
very rapid decrease in the magnitude of the eigenvalues. Notably,
we found that more of the spectrum fell within the range of the
Wishart distribution when zero-padding and zero-order hold was
practised (Figure \ref{fig:values}).

Analogous to the RMT approach, this paper compares the estimated
covariance matrices of Wilcox and Gebbie \cite{WiGe2} by
inspecting signal and noise content of the corresponding
eigenmodes (principle components). We consider the extent to which
properties of the derived time-series deviate from a Gaussian
null-hypothesis as a means of differentiating potential signal
from noise. Towards this end, we investigate
serial correlations, calendar effects and long-term memory in the
data.

\section{The Data and Eigenmodes}


\subsection{Correlation Matrices and Eigenmode Time Series}


The 10 years of data  were windowed to create 6 overlapping 5-year
subsets of daily price data consisting of $253, \ 293, \ 321,\ $
$330, \ 335$ and $341$ shares, respectively, ranging 1993-1997 to
1998-2002. Each block was screened to remove shares which were
de-listed or which traded very infrequently \cite{WiGe1}.
Investigating the effect of different treatments of measurements
for prices, our approach favoured the notions that (1) no trading
implies no price measurement, and (2) share cross-correlations can
only be computed when there are pair-wise measurements on the same
day.

We find the returns $r_i(t) = \ln\,S_i(t+\triangle t)/S_i(t), $
where $S_i(t)$ denotes the price of asset $ i \in \{ 1, \ldots ,
N\}$ at time $t$, and consider the cross-correlations in two ways.

\begin{description}
\item {\em The usual cross-correlation matrix}
is applicable to idealized data with non-zero price fluctuations
and no missing data: $ \ C_{ij} = ( \langle r_ir_j\rangle -
\langle r_i\rangle\langle r_i\rangle  )/ \sigma_i\sigma_j\,$,
where $ \langle \ldots \rangle $ denotes average over period
studied and $ \sigma_i ^2:= \langle r_i^2\rangle -  \langle
r_i\rangle ^2 $ is the variance of the price changes of
 asset $i$.
\item {\em The pairwise measured-data cross-correlation matrix} is applied
 when there is missing data in returns time series:
$\mathcal{C}_{ij}  =   (\langle \eta_i \eta_j\rangle - \langle
\eta_i\rangle\langle \eta_i\rangle)/ \varsigma_i\varsigma_j\,, $
where $\eta_i$ and $\eta_j $ denote subseries of $r_i$ and $r_j$
such that for each $i-j$ pair there exists measured data for both
$\eta_i$ and $\eta_j $ at every time period in the subseries, and
$ \varsigma_i ^2:= \langle \eta_i^2\rangle -
 \langle \eta_i\rangle ^2 $ (pairwise deletion method).
\end{description}
We compute correlation matrices for the 6 subsets of data in two
different ways to  address the problem of missing data and no
trading, i.e. price data with zero fluctuations for several day in
succession \footnote{In \cite{WiGe1} we also computed correlation
matrices without the zero-padding practise but with zero-order
hold interpolation.}.
\begin{itemize}
\item[A:]
We assign the value of zero whenever there is no measured data and
compute the correlation matrix in the natural way (zero-padding).
\item[B:] In the event of $2$ or more successive zero-valued
price fluctuations we delete the measured value $r_i(t)$. This
effectively turns the zero-valued returns into missing data. We
then compute the   measured-data correlation matrix (no zero-order
hold, no zero-padding).
\end{itemize}

\begin{figure}
  \centering
  \includegraphics[width=10cm]{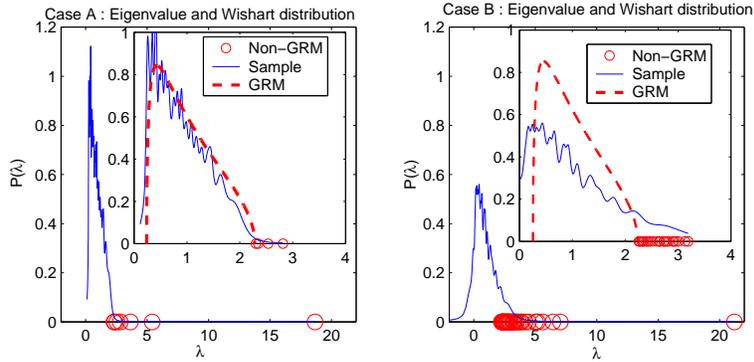}
  \caption{Daily price returns for JSE main board shares for years 1998-2002 were used to investigate
  eigenvalue structures of the  estimated correlation matrices. Here we
  show the eigenvalue density functions for cases A and B. Circles highlight
  distinct eigenvalues greater than the maximum RMT predicted value $\lambda_{\max}$ for the Q-factor of the sample.
  Insets: plots of the Wishart distribution (Eqn.~\ref{wishart}) are superimposed on plots of the small eigenvalues.
  }\label{fig:values}
\end{figure}

For each correlation matrix we order the eigenvalues from largest
to smallest and derive a corresponding time series as follows:

\bea x^{(k)}(t) = \sum\limits_{i=1}^{N}\,u_i^{(k)}r_i(t),
\label{emodes} \eea where $u^{(k)}$ denotes the eigenvector
corresponding to the $k$-th eigenvalue and $u_i^{(k)}$ are its
components. We refer to these as {\em eigenmode timeseries} or
simply {\em eigenmodes}. For Case A, there are no missing returns. Consistent with our construction for Case B, valid measurements of returns
exclude repeated returns when there was in fact no trading,  and such missing
returns are excluded in Equation (\ref{emodes}). Only if all returns for all stocks are
missing on a given day will there be no measured return for a given eigenmode.
Detailed discussion of the data filtering and partitioning is presented in \cite{WiGe2}.

\subsection{RMT Predictions for Behaviour of Eigenvalues: the Wishart
Distribution}

Let $M$ denote an $N\times L$ matrix whose entries are i.i.d
random variables which are normally distributed with zero mean and
unit variance. As $N,L \rightarrow \infty$ and while $Q= L/N$ is
kept fixed, the probability density function for the eigenvalues
of the Wishart matrix  $R = \frac{1}{L}MM^T$ is given by
\cite{bai,edelman,mehta2,mehta,pre60a}:

\bea p(\lambda) & = & \frac{Q}{2\pi} \frac{\sqrt{(\lambda_{\max}
-\lambda) (\lambda-\lambda_{\min})}}{\lambda} \label{wishart} \eea
for $\lambda$ such that $ \lambda_{\min}  \leq   \lambda \leq
\lambda_{\max}$, where $ \lambda_{\min}$ and $ \lambda_{\max}$
satisfy $\lambda_{\max / \min} =  1+\frac{1}{Q} \pm 2\sqrt{1/Q}.$

Figure ~\ref{fig:values} depicts  the eigenvalue density functions
for Cases A and B with plots of the Wishart distribution
(Eqn.~\ref{wishart}) superimposed. In this study, the length of time series used is $ L \approx$ 1305 and the number of stocks $N$ used to obtain the covariance matrix is 341 for Case A and 319 for Case B for the period 1998-2002.

For Case A, only 7 eigenvalues (less than 1\%) fell above the Wishart range. For Case B we found that $\approx$ 12\% of the eigenvalues fell above
$\lambda_{\max}=2.23$, a significantly larger percentage than  $\approx$ 6\% for the RMT analysis of daily price data for 406 stocks in the S\&P500 for 1991-1996 \cite{prl83a, prl83b, pre65a}. For both covariance matrix estimations we found  the largest eigenvalue to  be $\approx$ 9.5 times larger than $\lambda_{max}$ (compared to $\approx$ 25 times  for the S\&P500 study). While the null-hypothesis of Gaussian returns may be useful for identifying how zero-padding and zero-order hold add noise to the data, it is possible that the noise band $ [\lambda_{\min},\lambda_{\max}]$ for an appropriate null-hypothesis for SA market data is wider than suggested by Equation (\ref{wishart}).

\section{Periodicity, Serial Correlation and Long Memory}

Following Bachelier's groundbreaking work in 1900
\cite{bachelier}, intermittent empirical investigations of the
random walk hypothesis for stock returns gained momentum with the
advent of computers (cf. \cite{cootner2, cootner, kendall,osborne} and
references therein) and culminated in Fama's
definitive contribution on the efficient market hypothesis (EMH)
\cite{fama}. Today there is a vast body of literature on the three
forms of the EMH, including investigations for emerging markets and
South Africa in particular \cite{JeSm,SmRo}.

The classic work of Granger
\cite{Granger} discusses the implication of long range
periodicities in economic variables on the control of economies.
It was observed that cycles in price fluctuations are not strictly
periodic and, moreover, high-frequency noise and seasonal effects
may cloud signal. Subsequently, asymptotic decay of the
autocorrelation function  as a power-law $\rho(\tau) \sim \tau^{-\alpha}$ has
become widely accepted as a definition for {\em long memory}.

We first
compute the autocorrelation and power spectra of the derived
time-series to
compare  periodic, aperiodic and scaling signatures in eigenmodes
within and outside the Wishart range.
We then apply the variance ratio test of Lo and MacKinlay \cite{LoMac1} to test the
random walk hypothesis ({\it weak form} of EMH). We find that the
leading eigenmode exhibits significant short range dependency,
while eigenmodes corresponding to the noise range of the eigenvalue
spectrum display negative serial correlation. The
interpretation of the variance ratio statistic can be understood
in terms of a combination of autocorrelations, for example,
$\rho(1) = \mathrm{VR(2)}-1$ \cite{LoMac1}.

Long-memory is also characterised in terms of the Hurst
exponent\footnote{The Hurst exponent $H$ is simply related to the
exponent $\alpha$ of the autocorrelation function: $ H = 1 -
\frac{\alpha}{2}.$} $H$ obtained from the {\em rescaled-range}
analysis introduced by Hurst \cite{hurst} and refined by
Mandelbrot and Wallis \cite{MaWa1} and Lo \cite{Lo91}.
 We compute  Hurst exponents and apply the more recently
developed method of detrended fluctuation analysis (DFA)
\cite{HuIvChCaSt1,KaBuReHaBu1,LiGoCiMePeSt1,PeBuHaSiStGo1} to
compare scaling signatures in eigenmodes
within and outside the Wishart range.

The long-memory tests which we apply are heuristic but strictly non-parametric  estimators which exploit the scaling properties of long memory processes.
There is no single definitive long memory test; all known tests have limitations and typically a set of tests may be used to inspect data. The literature on this topic is extensive and some reviews of key aspects may be found in
\cite{alfarano,BrLi,LiFa,TaTeWi, WiTaTe}. Recently it has been shown that Markov processes, which by definition cannot have long memory, can have Hurst exponents $H\neq \frac{1}{2}$ \cite{BaGuMc}.

\subsection{Auto-correlation and Power Spectra of Eigenmodes }

Spectral methods are useful as heuristic diagnostic tools for
detecting periodicity and long-memory \cite{Beran,Granger}.

%


The auto-correlation of an eigenmode time-series  $x(t)$ is
computed as follows:

\beq \rho(\tau) =  \frac{1}{N-\tau} \sum\limits_{t=1}^{N-\tau}
z(t)z(t+\tau), \eeq

where $z(t) = \frac{x(t) - \bar{x}}{\sigma_x}$ , $\bar{x} =
\frac{1}{N}\sum\limits_{t=1}^{N} x(t)$ and ${\sigma_x} = \left[
\frac{1}{N-1}\sum\limits_{t=1}^{N}(x(t)-\bar{x})^2\right]^{\frac{1}{2}}.
$

Clearly $\rho(\tau) = 0 $ when the $z(t)$'s are uncorrelated for
all $\tau>0$, while if there exist short range correlations then
$\rho(\tau)$ decays exponentially. The presence of long-range
correlations gives rise to power-law decay $\rho(\tau)\sim
\tau^{-\alpha}$.

Determining the power spectrum is an established
practice for detecting correlation behavior. The periodogram gives
an estimate \footnote{By the Wiener-Khintchine theorem.}:

\beq S(\omega) = \frac{1}{ N}\left| \sum\limits_{t=1}^{N} z(t)e^{i
\omega t }\right|^2.\eeq

The spectral density of a stationary series with long-range
dependence exhibits power-law decay as $S(\omega)\sim
|\omega|^{-\beta}$.

\begin{figure}
\centering
\includegraphics[width=7.7 cm]{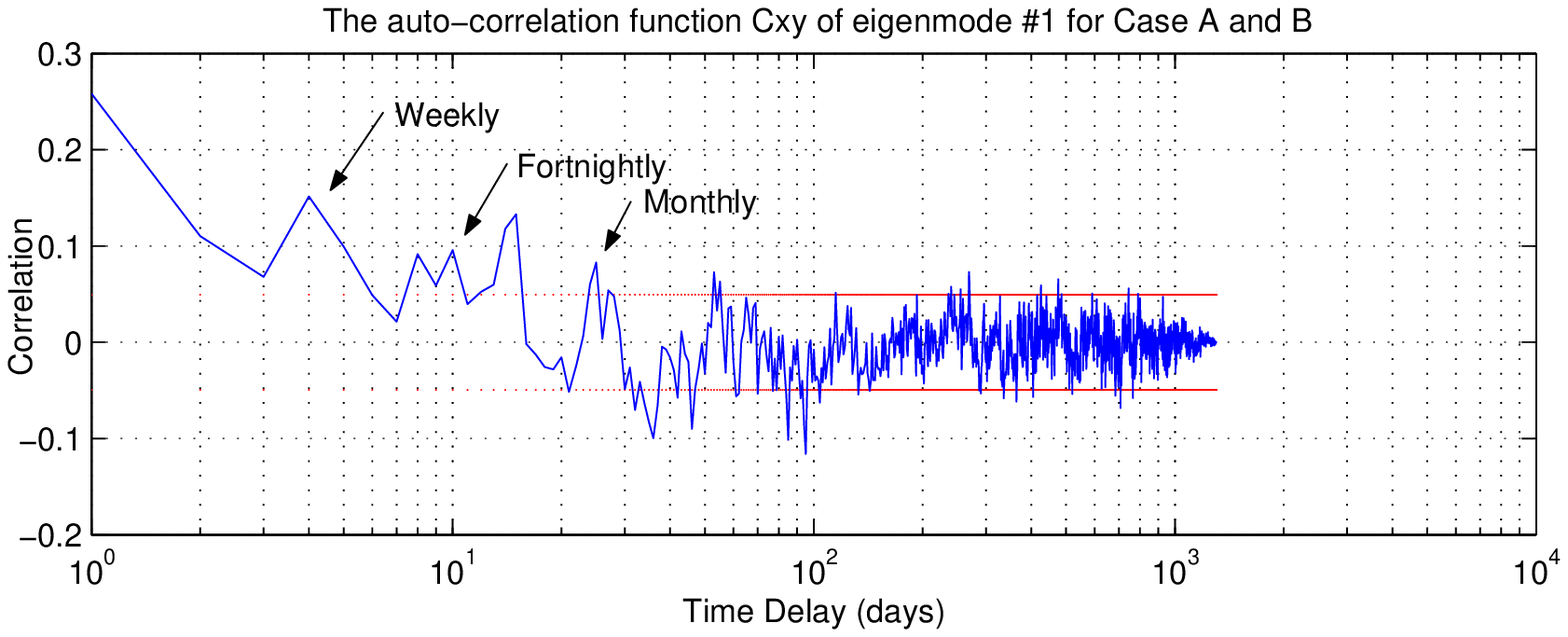}
\includegraphics[width=8 cm]{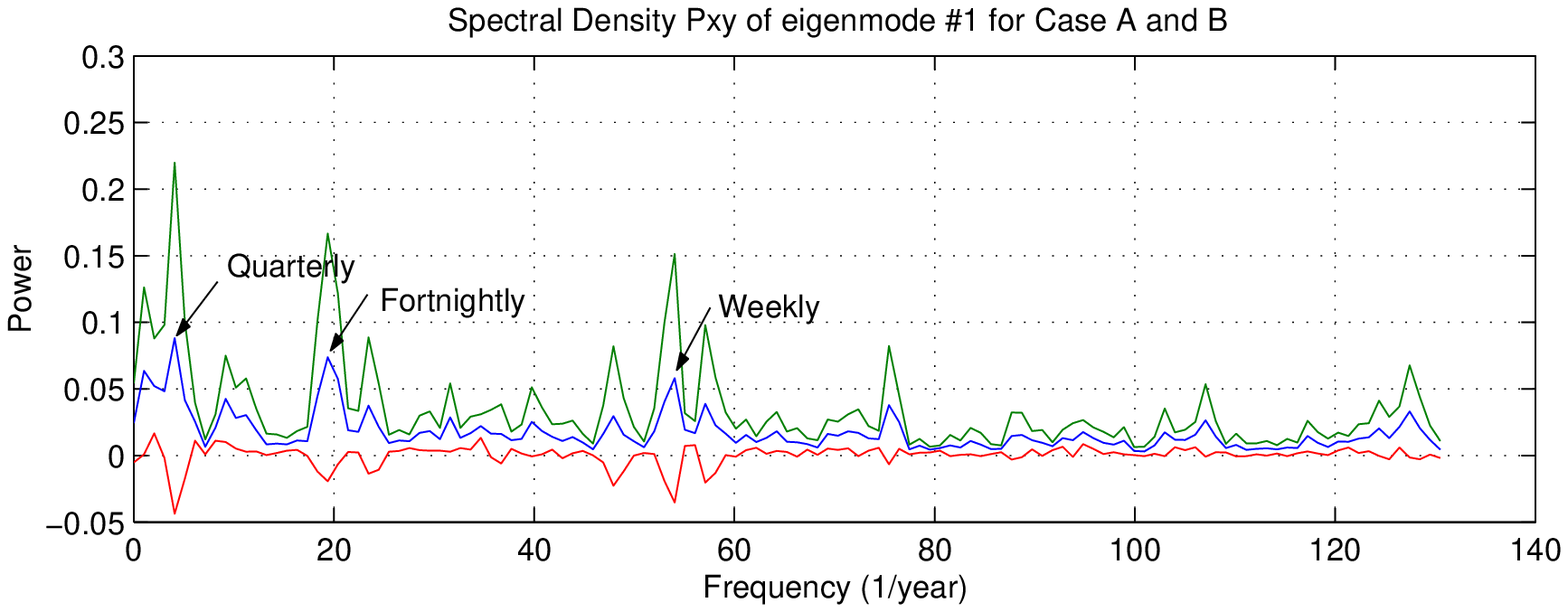}
  \caption{The auto-correlation function and power
  spectrum for the leading eigenmode for Case A are plotted; plots for Case B are
  (almost) identical to that of Case A.
   The autocorrelation functions (in both cases) have peaks near to 4 days, 8 days, 10 days,
  15 days, and 24 days.  Note that the zero
  lag point with auto-correlation 1 is not included on the graph.
  The power spectrum is plotted (middle of the three curves) with standard confidence intervals (top and bottom curve). The power
  spectra (in both cases)
  have peaks at 1, 4, 10-12, 20-24, 54-58 cycles per year; the three largest peaks are
  quarterly (4 cycles per year), monthly (20 cycles per year), and weekly
  (54 cycles per year). }\label{fig:acor1}
\end{figure}

\begin{figure}
\includegraphics[width=8cm]{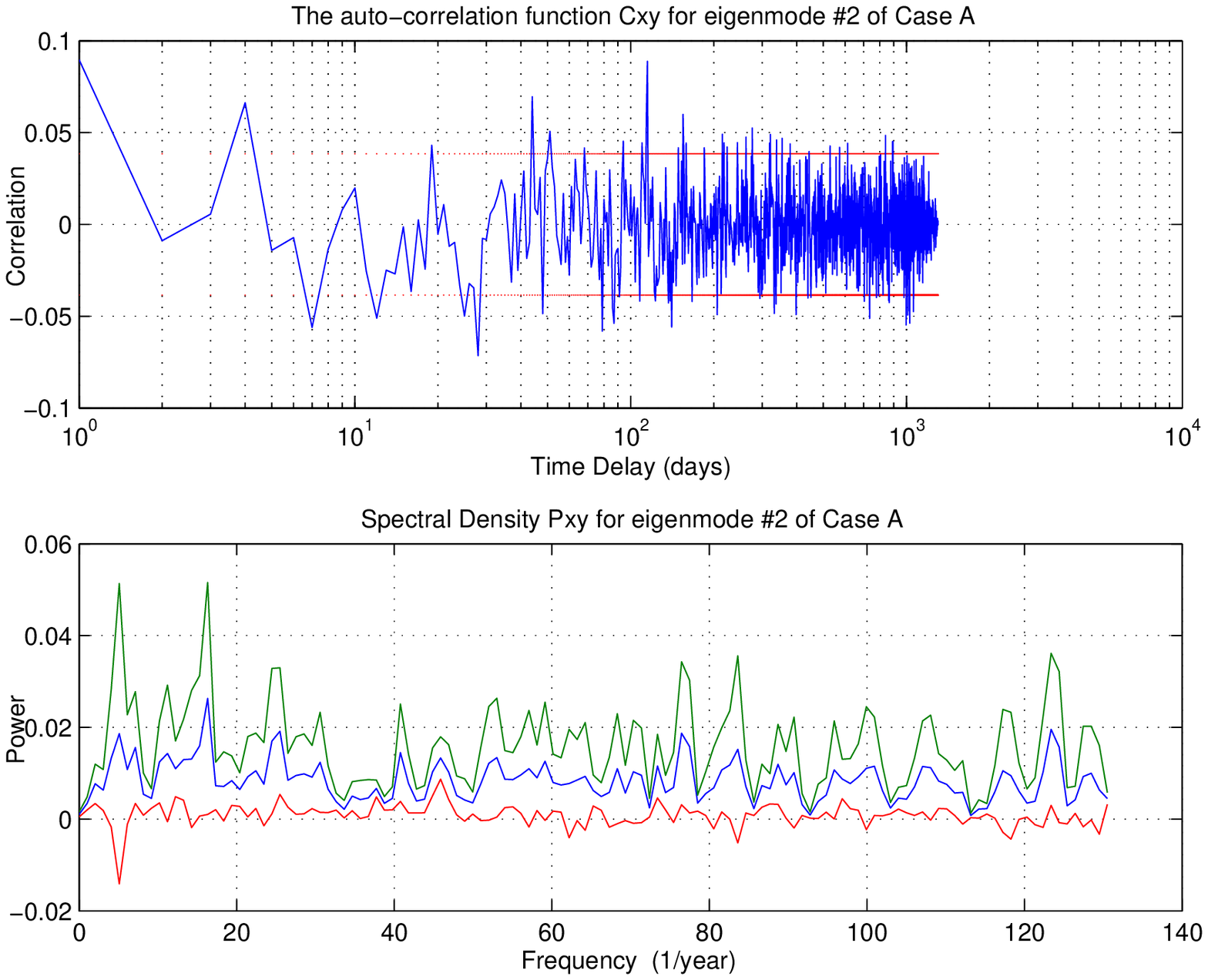}
\includegraphics[width=8cm]{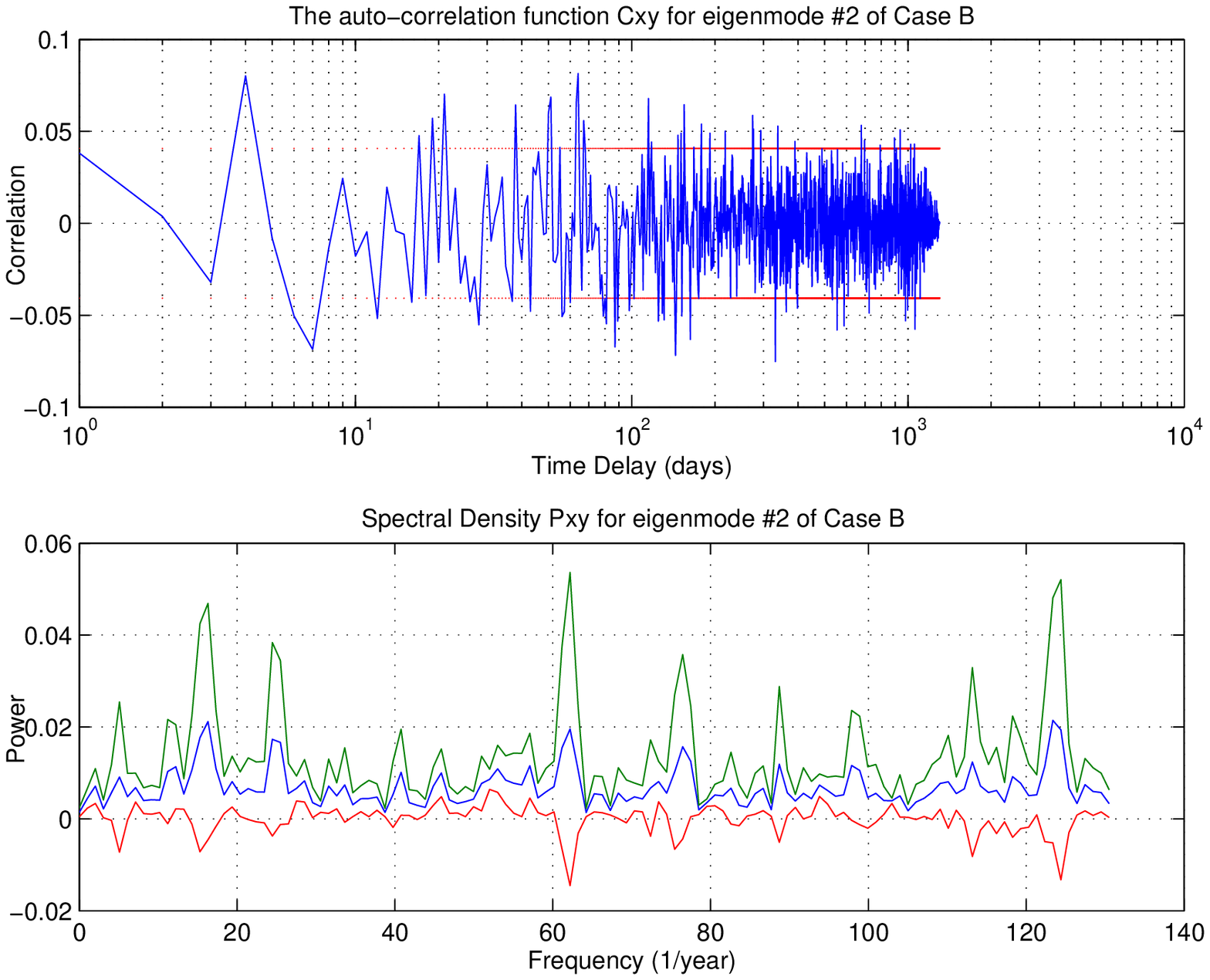}
  \caption{The autocorrelation functions and power spectra for the second eigenmodes
  for Case A and Case B are given. For both cases the auto-correlations
  and power are significantly less than those of the leading eigenmode (where quarterly peaks at 4 cycles per year
  attained values of $\approx$ 0.08 in the power spectrum).
  Weekly auto-correlations still exist for both cases. However, in Case A high-frequency noise
  effects begin to appear. Case B
  has a power spectrum with more discernable peaks (middle of the three curves with standard confidence intervals above and below). In particular
  there is still a peak for quarterly effects. }\label{fig:acor2}
\end{figure}

Figure \ref{fig:acor1} depicts the auto-correlation and power
spectrum for the leading eigenmode time-series (which corresponds
to the largest eigenvalue) for Case A; the functions for Case B
are almost identical. Figure \ref{fig:acor1} highlights
well-defined seasonality in the leading eigenmodes for both cases.
In particular, the auto-correlation functions have peaks which
suggest weekly, fortnightly, three-weekly and monthly periodic
calendar effects. There are peaks in the power spectra at: 1 cycle
per year (annual), 4 cycles per year (quarterly), 12 cycles per
year (monthly), 20-24 cycles per year (three-weekly to
fortnightly), 54-58 cycles per year (weekly); three largest peaks
are quarterly (4), three-weekly (20), and weekly (54).

The remaining eigenmodes for case A and case B are quite
different.  The auto-correlation functions and power spectra for
the second  eigenmodes are given in Figure \ref{fig:acor2}.
The distinct calendar effects witnessed for the leading eigenmodes
are less prevalent in these plots and those of the remaining
eigenmodes. Moreover, the calendar effects diminish more rapidly
for Case A than for Case B.

For Case A, aside from the eigenmodes corresponding
to the  3 largest eigenvalues, we found little evidence of
periodic behaviour. As the eigenvalues approach the
Wishart range, the power spectra become flat. Within the Wishart
 range and towards to smallest eigenvalues, the power spectra of
 corresponding eigenmodes exhibit more power at
 higher frequencies and very little power at the lowest frequencies (as
one would expect with randomly tampering with the data).

In Case B we find that there are seasonal effects in the
eigenmodes of the eigenvalues up to the upper bound of the Wishart
range. Within the Wishart range the power spectra of corresponding
eigenmodes are sufficiently flat to suggest no periodic
components. However, eigenmodes of the smallest eigenvalues once
again display seasonal (periodic) effects, with increases of power
for weekly and fortnightly cycles.

Higher-frequency trade-by-trade data is required for insight into
the behaviour of price fluctuation eigenmodes within periods of
less than 2 days (130 cycles per year).

Suggestions of dual peaks near the significant calendar effect
frequencies (e.g. fortnightly and quarterly frequencies) hint at
timescale mixing. This may be due to a combination of holidays and
missing data because the study uses the 261 day year (the work day
year).

The auto-correlations of the absolute values  of the eigenmodes,
i.e. $ \langle |z(t)|,|z(t+\tau)|\rangle , $ indicated strong
signatures of volatility clustering. There were suggestions of
peaks at fortnightly and quarterly intervals in the
auto-correlation function with a linear decline that reached the
noise floor after 60-100 days. In the power spectrum there were
signatures of annual, quarterly and fortnightly effects.

The correlations of eigenmodes with the absolute value of the
eigenmodes, i.e. $ \langle z(t),|z(t+\tau)|\rangle , $ were flat
and exhibited no discernable peaks.

\subsection{Variance Ratios for the Eigenmodes}

The variance ratio test of Lo and MacKinlay \cite{LoMac1} is based
on  the property that the sum of the variances is the variance of
the sum for i.i.d increments. Starting with a simple null hypothesis,
suppose that a price process follows geometric Brownian motion:
\begin{equation} dS_t =  \mu Sdt + \sigma S dW_t, \end{equation}
where $W_t$ denotes a standard Wiener process. In this case, the
discretely sampled returns, \linebreak $r(t) = \ln\,S(t )/S(t -
\triangle t), $ are uncorrelated and normally distributed. Letting
$r^q(t) $ denote the $q$-lagged returns, $ \ln\,S(t )/S(t -
q\triangle t) = r(t) + r(t-1) + ...+ r(t-q),$ we have $\textrm{
Var}[r^q(t)] = q\textrm{ Var}[r(t)]$, where $q \in \mathbb{Z}^+$.
Under this null hypothesis, the variance ratio, $\mathrm{VR}(q)$,
defined by
\begin{equation}
    \mathrm{VR}(q) = \frac{\frac{1}{q}\textrm{Var}[r^q(t)]}{\textrm{Var}[r(t)]},
\end{equation}
satisfies $\mathrm{VR}(q)=1$ for all $q$.

An interpretation of the ratio $\mathrm{VR}(q)$  follows from its asymptotic
behaviour \cite{LoMac1}:
\begin{equation}
\mathrm{VR}(q) - 1 \sim \sum_{j=1}^{q-1} \frac{2(q-j)}{q} \hat
\rho(j), \label{eqn:vr-intuition}
\end{equation}
in terms of the lag-$j$ auto-correlation estimate
$\hat \rho(j)$ for the return series. It can be seen that for $\hat \rho(j)= 0$,
$\mathrm{VR}(q) \sim 1$; similarly if $\hat \rho(j) = 1$ for all lags, then  $\mathrm{VR}(q) \sim q$, and for constant positive (negative) autocorrelation, we see that $\mathrm{VR}(q)$ increases (decreases) linearly with $q$. More generally $\mathrm{VR}(q)-1$ is a linear combination of the first $q-1$ autocorrelations
with arithmetically decreasing weights.

%
%

Lo and MacKinlay derive a more general test statistic for the null
hypothesis of uncorrelated heteroscedastic increments\footnote{See
Lo and MacKinlay for the precise conditions. The null hypothesis
allows for general forms of heteroscedasticity, including
deterministic changes in variance and ARCH processes.}, denoted
$Z^*(q)$:
\begin{equation}
Z^*(q) = \frac{\mathrm{VR}(q) -1}{\sqrt{\hat{\theta}(q)} },
\end{equation}
where $\hat \theta$ is the asymptotic estimated variance of the
$\mathrm{VR}(q)$
\begin{equation}
\hat{\theta}(q) =
4\sum\limits_{j=1}^{q-1}(1-\frac{j}{q})^2\,\hat{\delta}(j)\label{eqn:theta}
\end{equation}
and $\hat \delta$ is the asymptotic estimated variance of the
auto-correlation coefficient
\begin{equation}
\hat{\delta}(j) = \frac{\sum\limits_{k=j+1}^{nq} \left(r(k) - \langle r \rangle\right)^2 \left(r^j(k) - \langle r \rangle\right)^2 }{(\sum\limits_{k=1}^{nq} \left(r(k) - \langle r \rangle)^2\right)^2}.
\end{equation}

\begin{figure}
  \centering
  \includegraphics[width=15cm]{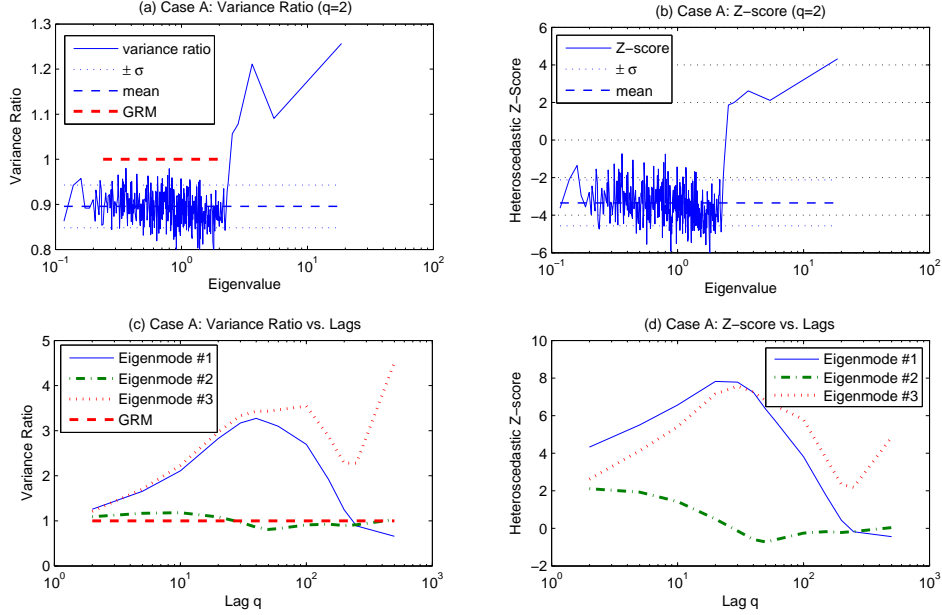}
  \caption{Daily price returns for years 1998-2002 were investigated using
  variance ratio tests at various lags for Case A. Moving left to right
  clockwise: graph (a) gives the q=2 variance ratio for all eigenvalues,
  the Wishart range is covered by the null case VR(2)=1, (b) provides the
  associated heteroscedastic Z-scores, (c) provides the variance ratio
  statistics for lags ranging from 2 through to 500 days, and (d) the
  associated heteroscedastic Z-scores for these.}\label{fig:vrA}
\end{figure}

\begin{figure}
  \centering
  \includegraphics[width=15cm]{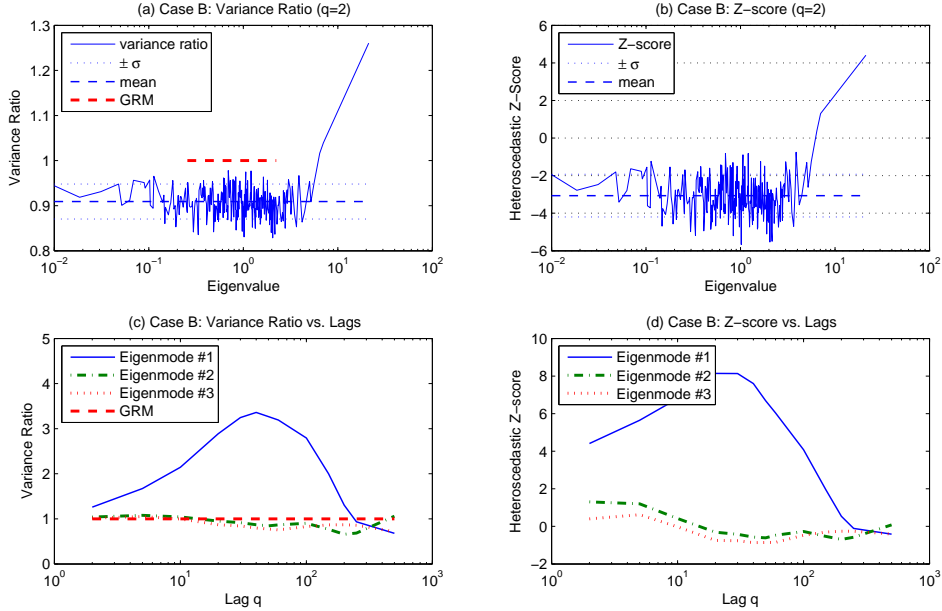}
  \caption{Daily price returns for years 1998-2002 were investigated using
  variance ratio tests at various lags for Case B as for Case A in Figure \ref{fig:vrA}.}\label{fig:vrB}
\end{figure}

Our investigation applies these tests heuristically. We note further that the original tests have limitations and there has been much subsequent work on improvements. Two significant contributions in the literature are: (a) the extension to a joint testing procedure by Chow and Denning to overcome the original focus which concerned the hypothesis that $VR(q)=1$ for individual lag $q$ \cite{ChDe},  and (b) the non-parametric approach based on ranks or on signs of White  to overcome the deficiency that the original test is based on asymptotic behaviour in a Gaussian setting \cite{wright}. A new test combination of these improvements is presented in \cite{colletaz}.

We let $ Z_{\lambda}^*(q)$ denote the value of $ Z^*(q)$ for the
eigenmode corresponding to the $\lambda$-th eigenvalue, where
$\lambda=1$ corresponds to the largest eigenvalue. Similarly we
denote $\mathrm{VR}_{\lambda}(q)$ for $\mathrm{VR}(q)$
corresponding to the $\lambda$-th eigenvalue. Figures
\ref{fig:vrA} and \ref{fig:vrB} summarise our investigation.

Figures \ref{fig:vrA} and \ref{fig:vrB}  (a) plot $
\mathrm{VR}_{\lambda}(2)$ against the full spectrum of
eigenvalues. Figures \ref{fig:vrA} and \ref{fig:vrB} (b) plot the
associated $ Z_{\lambda}^*(q)$ for the full spectrum of
eigenvalues. Figures \ref{fig:vrA} and \ref{fig:vrB} (c) plot $
\mathrm{VR}_{\lambda}(q), \ \lambda = 1,2,3$ as a function of lag
$q$. Figures \ref{fig:vrA} and \ref{fig:vrB} (d) plot $
Z_{\lambda}^*(q), \ \lambda = 1,2,3$ as a function of lag $q$.

For $q=2$, in Case A the eigenmodes corresponding to the 4 largest
eigenvalues exhibit positive serial correlation, while the remaining
eigenmodes have slight negative autocorrelation. For Case B only
the leading eigenmode suggests significant serial
correlation for $q=2$.

Varying $q$, for Case A we see that the $1^{st}$ and $3^{rd}$ eigenmodes  exhibit
 serial correlation, with $ \mathrm{VR}_1(q)$
increasing almost linearly up to 3.2 at q=40 and
$\mathrm{VR}_3(q)$ increasing almost linearly up to 3.5 at q=100.
$\mathrm{VR}_1(q)$ for Case B  is approximately the same as $
\mathrm{VR}_1(q)$ for Case A, while $\mathrm{VR}_3(q) \approx 1$
for all $q$ in Case B. In both cases we have that $\mathrm{VR}_2(q) \approx
1$ for all $q$.

For eigenmodes in the Wishart range,
Table \ref{tab:vrAB} presents the means
$\frac{1}{\Lambda}\sum_{\lambda \in W}
\mathrm{VR}_{\lambda}(q)$ for
increasing values of $q$, where $W = \{\lambda | \lambda_{min}\leq
\lambda \leq \lambda_{min}\}$ and $\Lambda = |W|$. We found that the average variance ratios decrease from $\approx$ 0.9
as a function of lag, with values for Case A slightly lower than Case B. This suggests slight negative serial correlation for eigenmodes in the Wishart range.

\begin{table}
\caption{\label{tab:vrAB} Case A and Case B mean values of
$\mathrm{VR}_{\lambda}(q)$ for eigenmodes corresponding to
eigenvalues, $\lambda$, in the Wishart range, $\lambda \in W$, for
increasing values of $q$ for the period 1998-2002.}
\begin{center}
\begin{tabular}{ccccccc}
  \hline
    \multicolumn{2}{c}{ \ \ \ \ }  & \multicolumn{2}{c}{ Case A}  & \ \ \ & \multicolumn{2}{c}{ Case B}  \\ & & & & \\
  q   & & $\ \frac{1}{\Lambda}\sum\limits_{\lambda \in W} \mathrm{VR}_{\lambda}(q)\ $ & $\ \sigma( \mathrm{VR}_{\lambda}(q))\ $ & & $\ \frac{1}{\Lambda}\sum\limits_{\lambda \in W} \mathrm{VR}_{\lambda}(q)\ $ & $\ \sigma( \mathrm{VR}_{\lambda}(q))\ $ \\ \hline
  2   & & 0.89 & 0.04 & & 0.91 & 0.03 \\
  5   & & 0.75 & 0.06 & & 0.76 & 0.06 \\
  10  & & 0.63 & 0.08 & & 0.65 & 0.07 \\
  20  & & 0.54 & 0.09 & & 0.56 & 0.08 \\
  30  & & 0.50 & 0.10 & & 0.51 & 0.09 \\
  40  & & 0.47 & 0.11 & & 0.49 & 0.10 \\
  50  & & 0.45 & 0.11 & & 0.47 & 0.11 \\
  60  & & 0.44 & 0.12 & & 0.46 & 0.12 \\
  100 & & 0.42 & 0.15 & & 0.45 & 0.15 \\
  150 & & 0.42 & 0.19 & & 0.46 & 0.19 \\
  \hline
\end{tabular}
\end{center}
\end{table}

%

We also inspected variance ratios for the large eigenmodes for the periods 1993-1997, 1994-1998, 1995-1999, 1996-2000 and 1997-2001. Notably, the $VR(q) \ vs. \  q$ plots for the periods 1996-2000 and 1997-2001 exposed  increasing variance ratio values\footnote{The rate of increase was nonlinear and possibly exponential.} for the first two eigenmodes for Case B. These periods coincide with the period following the August 1997 Russian GKO default\footnote{Some contextual facts for the SA market were reviewed in \cite{WiGe2}.}.

 Using weekly data, Lo and MacKinlay \cite{LoMac1} reported a variance ratio $\mathrm{VR}(2) = 1.30$ for equally weighted CRSP NYSE-AMEX index data over the period 1962 through 1985\footnote{Further analysis of this same data is discussed in \cite{WiTaTe}.}. In the context of South African data, Jefferis and Smith \cite{JeSm} found $\mathrm{VR}(2) = 1.01$ for  weekly ALSI 40 data for the period April 1996 through March 2001. Comparison with our study must take into account the differing data sampling frequencies and dissimilar universes of stocks.  We computed $\mathrm{VR}_1(2) \approx 1.26$ and $\mathrm{VR}_1(6) \approx 1.75$
 for the $1^{st}$ eigenmodes in both Case A and Case B (q=6 in Figures \ref{fig:vrA} and \ref{fig:vrB}). This suggests  positive auto-correlation over daily and weekly lag, which is corroborated by inspection of the auto-correlation function as given in Figure \ref{fig:acor1}. Jefferis and Smith
report variance ratio values of 1.21 and 1.33 for the mid capitalisation and small capitalisation
indices, respectively. Since our universe is biased towards mid- and small capitalisation
companies, we consider our results to be in broad agreement with the findings of \cite{JeSm}.

\subsection{Rescaled Range  and  Detrended Fluctuation Analysis (DFA) of
eigenmodes }

The {\em rescaled range} refers to the range over standard deviation
statistic developed by Hurst (1951) in an analysis of Nile
flooding \cite{Beran}. The now classical statistic of a
time-series is computed:

\beq \mathrm{H}(\tau) = \frac{1}{\hat{s}_{\tau}} \left[
\max\limits_{0\leq i \leq \tau} \sum\limits_{t=1}^{i} (x(t) -
\bar{x}_{\tau}) - \min\limits_{0\leq i \leq
\tau}\sum\limits_{t=1}^{i} (x(t) - \bar{x}_{\tau}) \right], \eeq

 where $\bar{x}_{\tau} =
\frac{1}{\tau}\sum\limits_{t=1}^{\tau} x(t),\ $
and $\hat{s}_{\tau}  = \left[
\frac{1}{\tau}\sum\limits_{t=1}^{\tau} (x(t) -
\bar{x}_{\tau})^2\right]^{\frac{1}{2}}$.

The asymptotic behaviour of $R/S$ for a stationary processes with
short-range memory should satisfy $E[R/S] \sim a\tau^{1/2}$. For a
long-memory process, $E[R/S] \sim a\tau^{H}$, where $H >
\frac{1}{2}$. The method is superior to spectral analysis which
detects periodic cycles and works well when records are long and
there are no trends. Lo introduced a modified $R/S$ statistic
\cite{Lo91} in order to discount the effect of short-range
dependance\footnote{Here we use the notation $\mathrm{H}^*(\tau)$ for the modification of $H(\tau)$ in the same sense that $Z^*(q)$ adjusts $VR(q)$ for heteroscedasticity. The weights $w_j(q)$ were proposed by
Newey and West for the estimation of the effect of serial
correlations in time-series data \cite{JDH94}.}:

\beq \mathrm{H}^*(\tau)   =  \frac{1}{\hat{\sigma}_{\tau}(q)} \left[
\max\limits_{0\leq i \leq \tau} \sum\limits_{t=1}^{i} (x(t) -
\bar{x}_{\tau}) - \min\limits_{0\leq i \leq
\tau}\sum\limits_{t=1}^{i} (x(t) - \bar{x}_{\tau}) \right]   \eeq

where \beq
 \hat{\sigma}_{\tau}^2(q)  =  \hat{s}^2_{\tau} + 2
 \sum\limits_{j=1}^{q}w_j(q)\gamma(j), \eeq
\beq w_j(q) = 1-\frac{j}{q+1}, \ \  q<\tau, \ \textrm{ and }  \
  \gamma(j) =
\frac{1}{\tau-j} \sum\limits_{t=1}^{\tau-j} x(t)x(t+j).\eeq

\begin{figure}
  \centering
  \includegraphics[width=15cm]{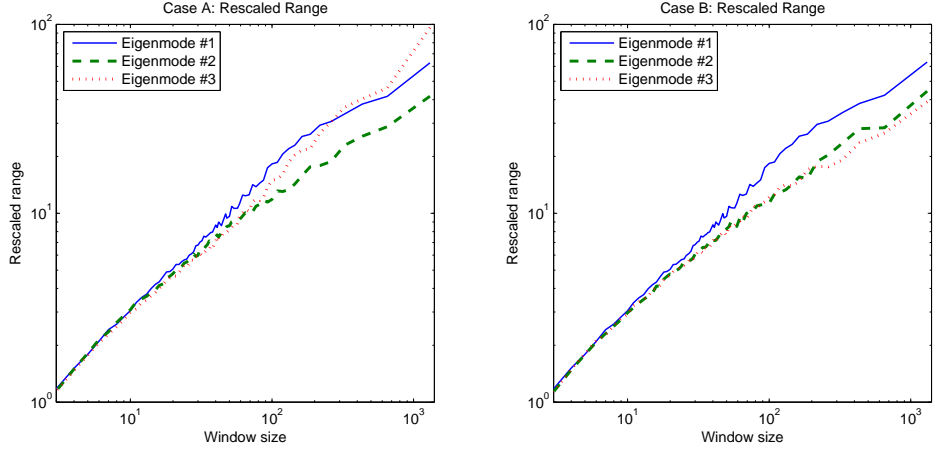}
  \caption{Modified R/S values as a function of box length for the 3 leading eigenmodes obtained from Case A and B covariance matrix estimations for daily price series of stocks listed on the JSE Mainboard 1998-2002.   }\label{fig:rs1}
\end{figure}

\begin{figure}
  \centering
  \includegraphics[width=15cm]{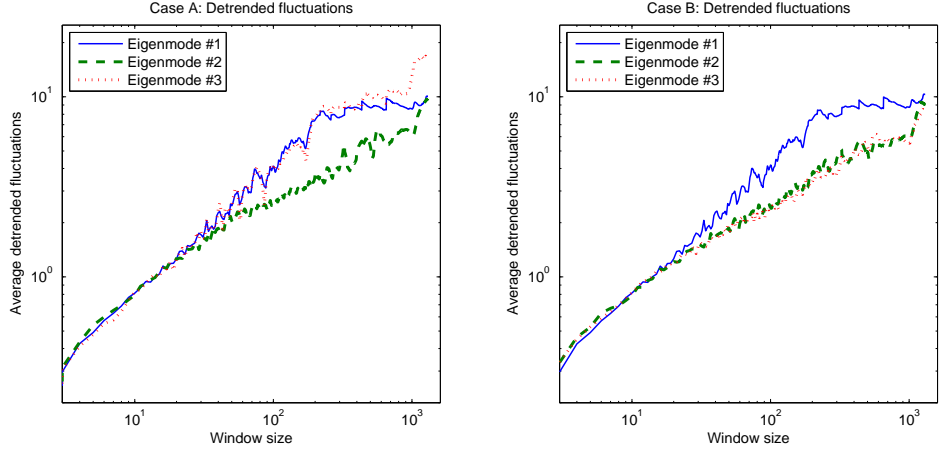}
  \caption{Detrended fluctuations as a function of box length for the 3 leading eigenmodes from Case A and B covariance matrix estimations for daily price series of stocks listed on the JSE Mainboard 1998-2002.  }\label{fig:df}
\end{figure}

\vspace{0.3cm}

DFA is a more recently developed method for detecting correlations
\cite{HuIvChCaSt1,KaBuReHaBu1,LiGoCiMePeSt1,PeBuHaSiStGo1}. It's
main strengths are the detection of long-memory in non-stationary
time series and the avoidance of false indications of long-memory
which are attributable to non-stationarity. The method is
particularly applicable when it is not known whether there are
underlying trends in the data or if the scales of underlying
trends are not known.
 The DFA algorithm is as follows : Integrate the time-series \footnote{Integration attempts to map the
time series to self-similar process.}

\beq y(t) = \sum\limits_{i=1}^t (x(t) - \bar{x}) \eeq

Divide the integrated time series into boxes of equal length
$\tau$. In each box fit a least squares line and detrend the
integrated time series by subtracting the local trend, denoted
$y_{\tau}(t)$, in each box. For each $\tau$, find root-mean-square
(r.m.s.) fluctuation of the detrended series :

\beq F(\tau) = \frac{1}{N}\sqrt{\sum\limits_{t=1}^N \left( y(t) -
y_{\tau}(t) \right) ^2 } \eeq

Clearly this quantity increases as the length of the subintervals
does. Moreover, it can be shown that $F(\tau) \sim \tau^{\gamma}$
for large $\tau$ \cite{KaBuReHaBu1}. If the detrended series is
uncorrelated or short-range dependent, then $F(\tau) \sim
\tau^{\frac{1}{2}}$. In particular, $\gamma = \frac{1}{2}$ implies
a lack of long range correlations, while $\gamma > \frac{1}{2}$
indicates an existence of long range persistence.



%

Figures \ref{fig:rs1} and  \ref{fig:df} compare the behaviours of the leading eigenmodes for Cases A and B. The R/S values for the $1^{st}$ eigenmode for Case A are almost identical to those for Case B. The distinction between Case A and Case B lies in the behaviour of the $2^{nd}$  and, more significantly, the  $3^{rd}$ eigenmode.  For Case A, inspection of the plots hint at crossovers in the scaling behaviour for the  $1^{st}$ and $3^{rd}$ eigenmodes. In Case B, the R/S plot for $3^{rd}$ eigenmode is very similar to that of the $2^{nd}$ and there is suggestion of crossover effects in all three leading eigenmodes.

   Just as for R/S values, the DFA values for the leading eigenmode for Case B are almost identical to those for Case A. For the DFA comparison,  distinction between Case A and Case B is more obvious. In Case B, the DFA plot for  $3^{rd}$ eigenmode is almost identical to that of the $2^{nd}$ (Figure \ref{fig:df}), while for Case A  the DFA plot for  $3^{rd}$ eigenmode is almost identical to that of the  $1^{st}$ eigenmode
(Figure \ref{fig:rs1}). In Case A, there is imperfect scaling for the  $1^{st}$ and  $3^{rd}$ eigenmodes. While inspection of the DFA graphs for the $2^{nd}$ and  $3^{rd}$ eigenmodes for Case B indicate that their fluctuations do exhibit scaling, we found that regressions suggest crossovers for all eigenmodes in both cases \cite{HuIvChCaSt1}. A more refined analysis is required for better interpretation of exponents computed from regressions.
    Such an investigation  would have to take into
account the known periodicities. Nevertheless, we tabulate exponent values in Table \ref{tab:exps}.

\begin{figure}
  \centering
    \includegraphics[width=15cm]{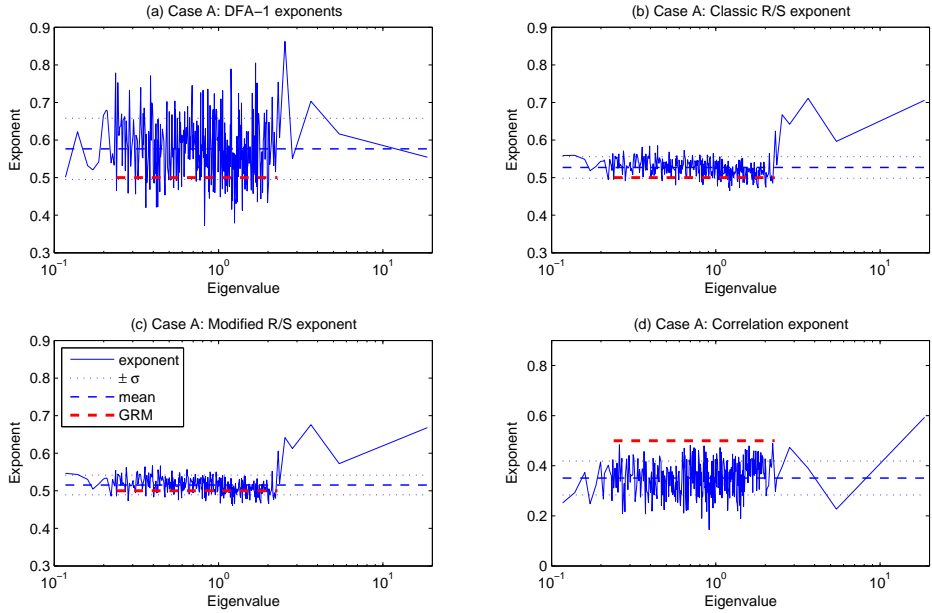}
  \caption{Values of DFA exponents, classic R/S exponents, modified R/S exponents,
  and the correlation exponents as a function of eigenvalues for eigenmodes from Case A. The expected Gaussian Random Matrix results are given
  within the Wishart range of eigenvalues. }\label{fig:scaling1}
\end{figure}

\begin{figure}
  \centering
  \includegraphics[width=15cm]{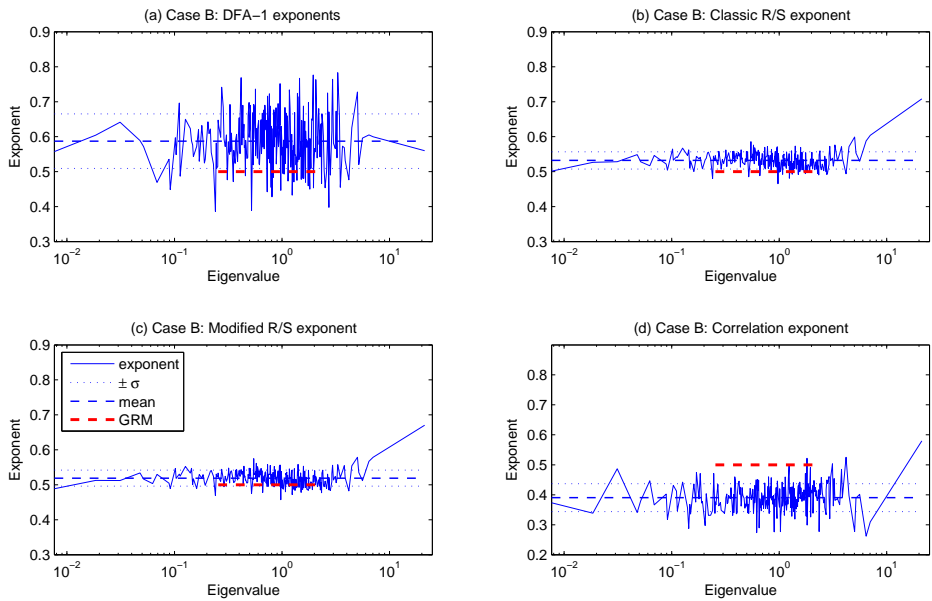}
  \caption{Values of the DFA exponents, classic R/S exponents, modified R/S exponents,
  and the correlation exponents as a function of eigenvalues for eigenmodes from Case B. The expected Gaussian Random Matrix results are given
  within the Wishart range of eigenvalues. }\label{fig:scaling2}
\end{figure}

   Figures \ref{fig:scaling1} and \ref{fig:scaling2} plot the exponents obtained from the four long-memory estimators applied to every eigenmode. This offers some comparison between properties  of the leading eigenmodes and those within the Wishart noise range.
   We generically let $E_{\lambda}$ denote an exponent computed for the eigenmode corresponding to the $\lambda$-th eigenvalue and include plots for
the  means $\bar{\mathrm{E}}_{\lambda} = \frac{1}{N}\sum_{\lambda}\mathrm{E}_{\lambda}$.
According to the mean DFA exponent values, there is weak persistence in price fluctuations for all eigenmodes,  with $\bar{\mathrm{E}}_{\lambda} \approx 0.58$ for Case A and $ \bar{\mathrm{E}}_{\lambda} \approx$ 0.59 for Case B. Estimates for exponents for the leading eigenmodes fall within the same band of values as those corresponding to eigenvalues in the noise range. It is possible that a more refined analysis to account for imperfect scaling and crossovers may yield lower DFA values.

According to the mean R/S exponent values, there is no significant indication of persistence for eigenmodes in the Wishart range. We find that   $\bar{\mathrm{E}}_{\lambda} \approx 0.53$  for classic R/S exponents for both Case A and B and   $\bar{\mathrm{E}}_{\lambda} \approx 0.52$ for  modified R/S exponents for Case A and B. The distinction between Case A and B lies in the R/S exponents for the leading eigenmodes which lie above the Wishart range.  For Case A the Hurst exponents suggest persistence in all 7 eigenmodes above the Wishart range, while for Case B there is suggestion of persistence for the $1^{st}, 2^{nd}$ and $4^{th}$ eigenmodes (out of 36 eigenmodes above the Wishart range for  Case B). Classic R/S exponents for the $1^{st}$ eigenmode are estimated to be 0.71 for both Case A and B, while R/S  exponent estimates for the $1^{st}$ eigenmode have slightly lower estimates of 0.67 for both Case A and B.
The higher values for estimates for the classic R/S exponents vs. modified R/S exponents may possibly be attributed to serial correlations, as uncovered via the variance ratio estimators discussed earlier.
We note also that it was found in \cite{TeTaWi} that the modified R/S statistic shows
a strong preference for accepting the null hypothesis of no long-range dependence even when
there is  long-range dependence  in the data.

For comparison, we include auto-correlation exponents in this section. Computations suggest persistence in the scaling behaviour of the $1^{st}$ eigenmode and anti-persistence for the rest of the spectrum. This is consistent with findings for the variance ratio investigations, particularly for Case B, where only the leading eigenmode had variance ratios significantly greater than 1. Nevertheless, finer analysis  to guage the validity of linear regressions for the autocorrelation exponent would be appropriate.

For all the long-memory estimators, the exponents for the $3^{rd}, 4^{th} $ and  $5^{th}$ eigenmodes are significantly higher for Case A compared to Case B. This confirms earlier observations about the $3^{rd}$ eigenmode based on Figures \ref{fig:rs1} and  \ref{fig:df}.
The dispersion of the exponents  is also higher for Case A  than for  Case B for all the
scaling exponents. These differences can be attributed to the noise introduced in Case A when zero-padding and zero-order hold was practised to treat missing data and illiquid trading.

\begin{table}
\caption{\label{tab:exps} Approximations for long-memory exponents of the  leading eigenmodes and  mean values means $\bar{\mathrm{E}}_{\lambda} = \frac{1}{N}\sum_{\lambda =1}^N
\mathrm{E}_{\lambda}$ for Case A and Case B covariance matrix estimations for the period 1998-2002.}
\begin{center}
\begin{tabular}{ccccccccc}
  \hline
      & \multicolumn{2}{c}{ DFA}  & \multicolumn{2}{c}{ Classic R/S} & \multicolumn{2}{c}{ Modified R/S}  & \multicolumn{2}{c}{ Autocorr}  \\
     & Case A & Case B  & Case A & Case B  & Case A & Case B  & Case A & Case B  \\ \hline
  $\mathrm{E}_{\lambda_1}$     & 0.55 & 0.56 & 0.71 & 0.71 & 0.67 & 0.67 & 0.59 & 0.58 \\
  $\mathrm{E}_{\lambda_2}$     & 0.61 & 0.61 & 0.60 & 0.60 & 0.57 & 0.58 & 0.23 & 0.31 \\
  $\mathrm{E}_{\lambda_3}$     & 0.70 & 0.60 & 0.71 & 0.53 & 0.68 & 0.51 & 0.47 & 0.26 \\
  $\mathrm{E}_{\lambda_4}$     & 0.55 & 0.52 & 0.64 & 0.54 & 0.61 & 0.53 & 0.30 & 0.40 \\
  $\mathrm{E}_{\lambda_5}$     & 0.86 & 0.73 & 0.67 & 0.60 & 0.64 & 0.58 & 0.50 & 0.39 \\
  $\bar{\mathrm{E}}_{\lambda \ }$ & 0.58 & 0.59 & 0.53 & 0.53 & 0.52 & 0.52 & 0.35 & 0.39 \\
  \hline
\end{tabular}
\end{center}
\end{table}

\section{Conclusions}

In this paper we examined serial correlations, periodic and
scaling properties of the eigenmodes derived from two different
methods of estimation of cross-correlations in South African financial market data from the JSE Mainboard the period 1993-2002. Data was partitioned into 6 five-year epochs and our report focuses on the last epoch 1998-2002. The first
estimation of cross-correlations, Case A, incorporated
zero-padding when there was missing data and zero-order-hold when
there were no price changes. For the second, Case B, correlation matrix
entries were found by restricting usual correlation calculations
to subseries of time-series pairs such that there were same-day
pair-wise measurements.

We  applied heuristic tests to eigenmode time series with an analysis of the market as a whole in view. More refined analysis, taking into account all known nuances for the data, would be required to make stronger inferences  about individual time series for financial application.
One would also need to review detailed features for all the eigenmodes in order to fully compare eigenmodes corresponding to eigenvalues in and outside of bounds of the Wishart distribution
given by random matrix theory predictions.

We show that the eigenmode fluctuations which correspond to the
large eigenvalues exhibit distinct calendar (periodic) effects and
that these effects are more pronounced for Case B.  Moreover, we
found that noise in the eigenmodes becomes more dominant as the
corresponding eigenvalues approach the Wishart range. In a previous study of
the same data it was shown that the bulk of the eigenvalues fell
within the Wishart range for Case A, and that about 88\% of eigenvalues
for Case B were in that range. In this study we found that
eigenmodes with significant periodicities correspond to
eigenvalues which lie outside of the Wishart range. Conversely, we
found that eigenmodes with high frequency noise and no discernable
periodic behaviour correspond to eigenvalues within the Wishart
range. This is consistent with inspection of the fundamental characteristics of the eigenmodes considered in \cite{WiGe2}, where it was reported that the characteristic compositions of eigenmodes are noticeably  varied outside the Wishart range for both cases, while they are almost identical within the Wishart range.

We inspected for serial correlations  using a  simple variance
ratio test. For eigenmodes in the noise band, we obtained declining variance ratio values from $\approx$ 0.9
 to just below 0.5 for increasing lags for both cases. Values for Case A were slightly lower than for Case B.
 The random walk hypothesis was rejected for the first eigenmodes in
both cases, and also for the second eigenmode for Case A. For the remaining large eigenmodes above the Wishart range, results could not rule out the random walk hypothesis. The spectral analysis and variance ratio findings suggest that interpolating missing data or illiquid trading days with zero-order hold introduces high frequency noise and spurious serial correlation for Case A. This inference is supported by the findings for the same time series presented in \cite{WiGe2}.

Long memory exponents identify deviations for eigenmodes corresponding to the five largest eigenvalues across both covariance estimators.
R/S analysis suggests long memory effects for all five eigenmodes for Case A and for eigenmodes one, two and five for Case B.  Unlike findings from the R/S analysis, DFA exponents did not clearly distinguish between large eigenmodes and those from the Wishart range and mean DFA exponents were comparitively high. On the other hand, mean autocorrelation exponents were comparatively low (significantly less than 0.5). We note
that, considering only estimated exponents results in loss of information, as was seen for the DFA where we discerned imperfect scaling and crossovers in the DFA plots (Figure \ref{fig:df}).

In this paper we do not offer explicit means to distinguish between the overall quality of information  for Cases A and B, as was done in \cite{WiGe2} where analysis clearly favoured Case B. We do highlight that properties of specific large eigenmodes are quite different for the two cases. This may be important for
portfolio optimizations where meaningful eigenmodes represent
systematic risk (risk factors collectively shared by the market at
large) that cannot be diversified away. It is the share-specific
residual modes that can be diversified away; if there are small
collective modes that have periodic behaviour or long memory, this could lead to
additional estimation uncertainties under portfolio optimization.

\section{Acknowledgements}

We thank Justin Solms for helpful conversations and the referee
for pertinent comments.
This research was partially supported by an South African NRF Thuthuka research grant.

\end{document}